\documentclass[pra,twocolumn,amsmath,amssymb, showpacs]{revtex4}
\usepackage{bm}

\usepackage{epsfig}

\begin{document}

\title{R. A. Fisher, Design Theory, and the Indian Connection}

\author{A.\  R.\  P.\  RAU \\
{\it Department of Physics and Astronomy, 
Louisiana State University, 
Baton Rouge, Louisiana 70803\\
(Fax, 1-225-578-6841;  Email, arau@phys.lsu.edu})}

\begin{abstract}
Design Theory, a branch of mathematics, was born out of the experimental statistics research of the population geneticist R. A. Fisher and of Indian mathematical statisticians in the 1930s. The field combines elements of combinatorics, finite projective geometries, Latin squares, and a variety of further mathematical structures, brought together in surprising ways. This essay will present these structures and ideas as well as how the field came together, in itself an interesting story.

\end{abstract}


\maketitle
    
\section{Introduction}

What do the following have in common:

$\bullet$  Kirkman's School Girl Problem: ``15 young ladies in a school walk out three abreast for 7 days in succession; it is required to arrange them daily, so that not two will walk twice abreast" \cite{ref1}.

$\bullet$  The puzzle-game SuDoKu, one of the greatest mathematicians, and judging effectiveness of fertilizers on potato varieties. 

$\bullet$  R. A. Fisher, statistician and population geneticist, a key figure in the synthesis between Darwin and Mendel.

$\bullet$  Branches of mathematics called design theory and coding theory. 

$\bullet$  Projective geometry, a subject in which unlike in Euclidean geometry, there is a duality between points and lines such that interchanging them in any theorem does not affect its validity.

$\bullet$ India's pioneering statistician and early associates in the school he founded. 

This essay will present the interesting mathematical structures and ideas in the above items and the human interest thread that weaves through them. Whether arranging numbers from 1 to 9 in a $9 \times 9$ array so that each numeral occurs once and only once in each row and column, arranging schoolgirls in $5 \times 3$ blocks so that no pair is repeated, or arranging plots of potato varieties and the laying of different fertilizers on them so that each variety is subjected to each type of fertilizer to gauge effectiveness, these are all problems of `experimental design' and now a branch of mathematics called `design theory' \cite{ref1,ref2}, related also to coding theory \cite{ref3}. These are parts of the wider fields of combinatorics as well as finite projective geometries \cite{ref4,ref5,ref6}. While some of the basics go back to the great mathematician Euler, it is the work of Fisher and of a school of Indian mathematical statisticians that gave birth to Design Theory \cite{ref7}. In statistics, this is also referred to as `Design of Experiments' or `Experimental Designs'.

\section{Design Theory}

Kirkman's School Girl Problem, originally posed by W. S. B. Woolhouse in 1844 \cite{ref8} and solved by the Rev. Thomas Kirkman, a Lancashire clergyman and amateur mathematician \cite{ref9}, in 1847 in a charmingly named journal \cite{ref10}, is a precursor of what have come to be known as `designs' and more specifically, `balanced incomplete block (BIB) designs' \cite{ref5,ref11,ref12} or `Steiner triple systems' \cite{ref5,ref13}. There was also early work by the great mathematician Euler and today, all of this is part of a branch of mathematics called design theory \cite{ref2}. 

The idea is to consider two sets, members of one to be allotted to those of the other with certain specified conditions. The first set of $v$ objects or symbols (may be anything: numbers, potatoes, \ldots), as with $v=15$ ladies, is to be put into $b$ blocks. Each block contains exactly $k$ distinct symbols, as in $k=3$ ladies abreast, each symbol to occur in exactly $r$ different blocks and every pair of distinct symbols to occur together in exactly $\lambda$ blocks. In the case of the school girls, $r=7$, the number of days, and $\lambda=1$ because no two should recur from one day to the next. Kirkman constructed the solution with $b=35$, these being the number of rows of three, 5 for each of the 7 days.
 
A $(v, b, r, k, \lambda)$ design or BIB is thus one of $v$ objects in $b$ blocks with each block containing exactly $k$ distinct objects, each object occurring in exactly $r$ different blocks and every pair ($t=2$ or more general $t$) of distinct objects occurring together in exactly $\lambda$ blocks. Block designs with $k=3$ are called triple systems. Those with $\lambda=1$ are called Steiner systems $S(t, k, v)$ and, if $k=3$ as well, Kirkman or Steiner triple systems $S(2, 3, v)$ because the Berlin mathematician Jakob Steiner, proposed their existence in 1853, conjecturing that the number $v$ had to be such that it would leave a remainder of 1 or 3 upon dividing by 6 \cite{ref14}. This was proved by Reiss \cite{ref15} six years later but they were unaware of Kirkman's work \cite{ref16}. 

The following relationships define a BIB: $vr=bk, \lambda (v-1)=r(k-1)$. For triple systems with $k=3$, these reduce to $ r=\lambda (v-1)/2, b=\lambda v(v-1)/6$. Another notation used for BIBs is $t-(v, k, \lambda)$ so that a Steiner triple system is $2-(v, 3, 1)$, the Kirkman problem being a 2-(15, 3, 1) design. An even smaller one is 2-(7, 3, 1) or $S(2, 3, 7)$ which we will encounter in Section 4 in a geometrical context of placing 7 points on 7 lines such that each line has three points on it and each point lies on three lines with no pair of points on more than one line. The terminology of symbols and blocks is replaced by the geometrical ones of points and lines, respectively. With $(v=b, r=k)$, such a BIB is said to be symmetrical. The result of Steiner and Reiss allows a parametrization of Steiner triple systems in terms of a single integer $n$. One family has $(v=6n+3, b=(3n+1)(2n+1), r=3n+1)$ and a second $(v=6n+1, b=n(6n+1), r=3n)$. With increasing $v$, establishing the exact number, 80 for $v=15$ and over two million for $v=19$, and classifying Steiner triple systems becomes complicated. For these and the long history of establishing the result of two non-equivalent designs for $v=13$, see \cite{ref17}.    

Much development of the subject comes from the work of R. A. Fisher who formulated the principles of statistical designs in 1925 in the context of agricultural research/statistics, and from Yates who introduced the use of BIB designs in 1936 \cite{ref11}. In studying the effects of various fertilizers and soils on growing potatoes and barley, Fisher was conducting field studies which led to the design of statistical experiments. A complete experiment on the effectiveness of $v$ different fertilizers on $b$ types of plants would require $b$ plots, each subdivided into $v$ areas. This could be prohibitively expensive. An `incomplete' one would test every type of plant with $k<v$ different fertilizers such that any two fertilizers would be tested on $\lambda$ different types of plants. `Balancing' occurrence of pairs of treatments on exactly $\lambda$ of $b$ blocks of size $k$ means the regular appearance of pairs of fertilizers on the same plant, allowing a complete covariance analysis of the results. This was Fisher's great insight along with his focus not on one character at a time but a multivariate analysis. He introduced the idea of variance and maximum likelihood, established inequalities named for him (that a proper BIB requires $b \geq v, r \geq k$), and rapidly in the 1920s and 1930s established the field with mathematical rigour, writing his 1935 book, ``The Design of Experiments" \cite{ref18}. The terminology introduced by Yates \cite{ref11} of v(arieties), t(reatments) and r(eplications) provides the symbols still in use today.

The next step in the development of Design Theory as the full-fledged branch of mathematics that it is today can be traced to Fisher's trip to India in 1938 when he visited his friend P. C. Mahalanobis who had similarly pioneered the use of agricultural statistics in India, establishing a journal, Sankhya, and the Indian Statistical Institute in December 1931. A couple of young assistants in that group, most notably R. C. Bose, with physics and mathematics background, had been following Fisher's idea of representing an $n$-sample by a point in $n$-dimensional Euclidean space, and were solving many design problems and constructing BIBs. They took up questions Fisher posed on statistical designs for controlled experiments, using their expertise in finite geometries, leading to the study based on Galois fields that forms the modern basis of the subject. We will return to this in Section V. 

\section{R. A. Fisher}

Ronald Aylmer Fisher, born in 1890, was a pioneer in mathematical statistics and made fundamental contributions to genetics, combining Mendelism with biometry. Other famous scientists of the time such as Bateson, Pearson and deVries saw conflicts between Mendel and Darwin, between the conserved, discrete types of the former and the small differences of continuous variation as the template for adaptive change in the latter's evolutionary theory. Already as an undergraduate in 1911, Fisher set this right by showing how indifferent variations could persist in a population even in a constant homogeneous environment. His 1930 ``The Genetical Theory of Natural Selection" was the first synthesis of these two pillars of modern biology \cite{ref19}. Other famous contributors such as Sewall Wright and Haldane soon followed in the early 1930s.

There are several biographies \cite{ref20,ref21,ref22,ref23}, including one by his daughter \cite{ref24}, of Fisher. Excerpts drawn from them, and other compilations that are presented here in this section, are meant only as a merest sketch to point the readers to more details in these sources. See also the website http://digital.library.adelaide.edu.au/coll/special/fisher/. In Cambridge in 1909, Fisher studied mathematics and physics (statistical mechanics) and also read Karl Pearson's ``Mathematical contribution to the theory of evolution". After four years (1915-1919) as a school teacher, Fisher joined the Rothamsted experimental station where Pearson had a group. Rejected for WW I because of poor eyesight, he took to farming as a eugenic way of life. In his field experiments, he developed the ideas of multivariate analysis and maximum likelihood and the block designs mentioned in Section II. 

Keeping statistical considerations in the planning and layout of experiments led to the `design of experiments'. Throughout his career, Fisher regarded statistical laws as basic and, interestingly, took his cue also from Heisenberg's contemporaneous work in quantum physics. He is also said to have commented that ``geometry had led to humanity's first great stage of intellectual liberation by discovering the principles of deduction, and that biometry was leading the second stage by discovering the principles of induction" \cite{ref23}.

The problem of design consists of choosing a set of treatments for comparison, specifying what varieties to which they are applied, randomizing rules for applying the treatments to the varieties, and specifying what is to be measured, the records then subjected to statistical analysis. Fisher's playful humour is apparent in some of the examples in his book \cite{ref18}. One deals with a lady of discernment who claims to be able to tell whether milk or tea was added first! If one were to present to her six cups of tea, three each mixed in each way, since there are 20 combinations of 3 out of 6 (given by $6 \times 5 \times 4/1 \times 2 \times 3$), there would be 1 chance in 20 of accidentally guessing the correct set. This 5\% is often taken as a standard level of significance and to do better, she should be given 8 cups with 4 of each preparation. Now a pure chance success reduces to 1 in 70 (the number of combinations of 4 out of 8 being $8 \times 7 \times 6 \times 5/1 \times 2 \times 3 \times 4$). Fisher goes on to discuss how to assess the significance of her discernment were she to get 3 correct and 1 wrong.

Another amusing example, with resemblance to Kirkman's schoolgirl one, proceeds thus in Fisher's presentation \cite{ref18}: 16 passengers on a liner discover that they are an `exceptionally representative body': 4 of them are English, 4 are Scots, 4 Irish and 4 Welsh. Further, they fall into four age groups, 4 being 35, 4 others 45, 4 more 55 and 4 being 65, with no two of the same age being of the same nationality. Next, it turns out that there are 4 lawyers, 4 soldiers, 4 doctors and 4 clergymen with again, the reader will get the picture, no two of the same profession sharing the same age or same nationality. It goes on, that 4 are bachelors, 4 married, 4 widowed and 4 divorced, with again no two of the same marital status sharing the same profession, age or nationality. Finally, the same with their political persuasion, 4 being conservatives, 4 liberals, 4 socialists and 4 fascists. With this somewhat head-reeling setup, Fisher poses that 3 among the fascists are known to be an unmarried English lawyer of 65, a married Scot soldier of 55 and a widowed Irish doctor of 45. It is easy enough to answer the first question of identifying the remaining fascist. Fisher's second question is to say that it is ``further given" that the Irish socialist is 35, the conservative of 45 is a Scot, and the Englishman of 55 is a clergyman, and then he asks what we know of the Welsh lawyer!      

Already in his undergraduate years at Cambridge, the subjects of evolution, the implication of Darwin for the human race, the results of Mendel, and Francis Galton's emphasis on selection continuously increasing the genetic inheritance of man, influenced him deeply in both basic and applied aspects. He formed the Cambridge Eugenics Society, while also working for twenty years with the Eugenics Education Society in London whose president was Leonard Darwin, the second youngest son of Charles Darwin. With genetics as the mechanism of inheritance and statistics as the correct tool for studying populations, the eugenic possibility of improving the biological inheritance of man was a theme in his thinking. In this he was an idealist, believing that eugenics societies must be involved in scientific research lest ``social scientists divert the Society from its proper study of human inheritance to serve a non-eugenic social function" \cite{ref24}. Against objections, he brought in several scientists from his Rothamsted association into the Society.

Today, in the post WW II world, the word eugenics is itself so discredited that it seems astonishing to see some of the references in the design literature, including many of Fisher's papers, in journals (now defunct) carrying that name \cite{ref12,ref25,ref26,ref27}. In 1933, Fisher took the Chair of Eugenics at University College, London, which housed the {\it Annals of Eugenics}, started in 1925 by Karl Pearson, the previous holder of that Chair. Fisher held that position and headed the Galton laboratory till 1943 when he moved to the Arthur Balfour Chair at Cambridge. He wanted to take the journal that he had fostered with him but University College kept it. An alternative he wanted was the {\it Journal of Genetics} but Haldane took that over, also in University College. As a result, Fisher started in 1947 the journal {\it Heredity}, now held by the Genetical Society of Great Britain. As can be seen by the references in this essay, many papers on design were published in these journals in the 1930-1950s. It should also be noted that {\it Annals of Eugenics} was originally designed to house eugenics and human genetics while the journal {\it Biometrika} would have papers in statistical methodology, but under Fisher, the former also became important for papers in statistics. 

Fisher had a long association with India, visiting it on many occasions over the decades, including the memorable one mentioned at the end of Section II. These will be taken up in Section V. Fisher spent his last years in Australia, dying in Adelaide in 1962.

\section{Finite Projective geometry, designs, and codes}

Most people are familiar with Euclidean geometry from school, with its axioms about points and lines and its propositions and proofs about triangles and circles. Two distinct points define a line and two lines either intersect at a point or are parallel. In the latter case, also familiar is the concept of points at infinity, two parallels regarded as meeting at infinity. Every child knows this as a matter of perspective, with parallel rail tracks a canonical example. Projective geometry \cite{ref28}, which removes the distinction between `finite' points and those at infinity, regarding all of them equally, is therefore important for perspective in art and architecture. 

Further, a distinguishing characteristic is that points and lines are on an equal footing with a `duality' between them unlike in ordinary Euclidean geometry. Thus, that two points define a line is in balance with two lines always meeting at a point, albeit one at infinity. A striking diagram, familiar in projective geometry, makes this clear, the two triangles (abc) and (ABC) `in perspective' with respect to the point P and with respect to the line (123) (Fig. 1). Like vertices of the triangles are connected by rays to P and like sides of the triangle, upon extension, meet on the common line (123). The triangles may lie on a plane or be arbitrarily oriented in space. If the two planes of the triangles are parallel, so that the extensions do not meet, the line (123) and its three points recede to infinity but the basic result remains. Fig. 1 is a partial Steiner system. 

\begin{figure}
\scalebox{2.}{\includegraphics[width=1.8in]{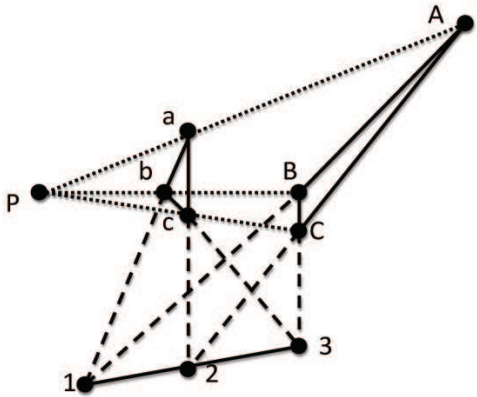}}
\vspace{-.2in}
\caption{The Desargues diagram of projective geometry. The two triangles, the rays from point P, and the line (123) constitute ten points on ten lines, with each point on three lines and vice versa. The two triangles are said to be in perspective with respect to P and to (123).}
\end{figure}

Finite geometries may be somewhat less familiar but first a couple of remarks about finite arithmetics, which again most are familiar with from the 12- or 24-hr clock. Technically called modular arithmetic, with a number such as this 12 the modulus, one deals only with the residues left over upon dividing by the modulus so that the only numbers that occur are less than it. The result noted in Section II about symmetric Steiner triple systems existing only for numbers that leave remainder of 1 or 3 upon division by 6 is an example, expressible as $v\equiv 1, 3$ (mod 6). 

Turning to finite geometries, instead of the continuous one familiar from school, one deals only with a finite number of points and lines. Thus, the finite Euclidean geometry with standard notation $EG (n, s)$ has $s^n$ points. One of the smallest, $EG (2, 2)$, has thus $2^2=4$ points. Correspondingly, the number of pairs out of four being six, there are 6 lines. Various diagrammatic representations are possible, one being a square with non-intersecting diagonals, but Fig. 2 shows a convenient one with the vertices of an equilateral triangle and its in-centre as the points. Using $(x,y)$ to represent a point, with $x$ and $y$ taking on only two values 0 and 1, the points can be denoted as shown. Some of the lines meet at a point, others do not. Thus, each side of the triangle and the line connecting the in-centre to the opposite vertex do not, and can be regarded as `parallel'. 

\begin{figure}
\vspace{-.2in}
\scalebox{2.0}{\includegraphics[width=1.6in]{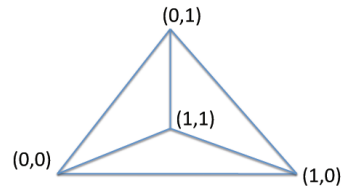}}
\vspace{-.5in}
\caption{The four points of a triangle and its centre, and the six lines shown form the Euclidean geometry $EG (2,2)$ \cite{ref6}.}
\end{figure}

Extending now to finite projective geometries \cite{ref5,ref29}, $PG (n, s)$, one adds to $EG (n, s)$ points and a line at infinity to restore the point-line symmetry/duality. Thus, with $EG (2, 2)$ in Fig. 2, imagine extending the lines from a vertex to the in-centre to meet the corresponding side of the triangle. With these two lines `parallel', the mid-point of the side where they meet is a point at infinity. Adding these three mid-points makes the total number of points 7. At the same time, the three points at infinity lie on a line, the in-circle, as shown in Fig. 3. There are then both 7 points and 7 lines in this diagram which indeed represents the finite projective geometry $PG (2, 2)$. In general, $PG (n, s)$ has $(s^{n+1} -1)/(s-1)$ points and for $PG (2, s)$, this number is $(s^2+s+1)$ points. 

In such a projective geometry, every pair of points lies on a unique line, and every line contains at least 3 points, one of them sometimes a point at infinity. Also, there is a set of 3 points not on a common line, an example being the vertices of the triangle in Fig. 3. $PG (2, 2)$ in Fig. 3 has a further property, that every pair of distinct lines contains a common point. Such an entity is called a `projective plane'. Fig. 3 is the smallest possible and is called `The Fano Plane' \cite{ref1}, arising in many varied contexts in basic and applied mathematics. In a projective plane, there exists what is called a `quadrilateral', that is, four points, no three of which lie on a line (the top four points ($e_1, e_4, e_6, e_7)$ in Fig. 3 provide an example). Its dual statement can be used as an alternative, that there exist four lines, no three of which go through the same point (the three sides and in-circle in Fig. 3 an example).

The assonance of the previous paragraph to items in designs in Section 2 must be evident. Indeed, the Kirkman design $(v=15, b=35, r=7, k=3, \lambda=1)$ is a $PG (3, 2)$ and the symmetric BIB or Steiner triple system $S(2, 3, 7)$ with $(v=b=7, r=k=3, \lambda=1)$ is $PG (2, 2)$. All that it takes to make the correspondence is to identify the symbols or objects of BIB with the points of projective geometry and, similarly, blocks with lines. In the alternative notation introduced earlier in Section II, these two geometries/designs are, respectively, $2- (15, 3, 1)$ and $2-(7, 3, 1)$. Another projective plane is $PG (2, 3)$ with 13 points and lines, and it is a $2-(13, 4, 1)$ design. It is a symmetric BIB with $(v=b=13, r=k=4, \lambda=1)$ but not a Steiner triple system because $r$ and $k$ are now 4, which represents the number of lines now on a given point.

\begin{figure}
\vspace{-.5in}
\hspace{-0.8in}
\scalebox{2.}{\includegraphics[width=2.in]{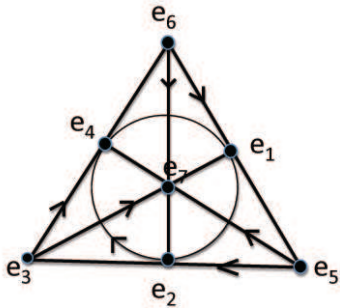}}
\vspace{-.4in}
\caption{The Fano Plane $PG (2, 2)$ of projective geometry with seven points and seven lines obtained upon adjoining in Fig. 2 three further points in the middle of the sides and a circle joining them. Also, the multiplication diagram for the seven octonions shown. The product of any two on a line equals the third with a $+/-$ depending on the direction of (along/against) the arrow \cite{ref39,ref40,ref42}.}
\end{figure} 

Given these intimate connections between designs and finite projective geometries, it is not surprising that Fisher and other pioneers to be considered in the next section, made fundamental contributions in both areas. Coding theory is another closely related subject, error correcting codes being important both in classical cryptography \cite{ref3} and today in quantum cryptography \cite{ref30,ref31}. See the Appendix for further remarks and connections to other areas of mathematics.     

\section{Connection to and contribution by Indian statisticians}

In India, at that time in the British Raj, P. C. (Prasanta Chandra) Mahalonobis started the serious study of agricultural statistics. Trained as a physicist, he pioneered statistics research in India, establishing the Indian Statistical Institute (ISI) and a journal Sankhya (in Sanskrit: number or determinate knowledge) in 1931, both of them respected institutions to this day. He had become a friend of Fisher and had visited him in Rothamsted in 1926-1927. Indeed, Fisher seems to have had a behind the scenes influence on the Government of India and the Indian Council for Agricultural Research (ICAR) in supporting Mahalanobis from 1927, and on Viceroy Linlithgow's support in establishing ISI. Sankhya was run out of private funds. A physicist S. S. Bose was hired as an assistant in 1929, working on problems of design, and two mathematicians, R. C. Bose and S. N. Roy, in 1931 \cite{ref24}. 

Mahalanobis felt that statistics was not supported by scientific and governmental authorities in India. He was brushed off by the Indian Science Congress when he asked for a section on statistics, the suggestion met with a scoff, that if statistics can be admitted, then why not astrology! Therefore, he arranged for a special Statistical Conference in Calcutta to follow the Indian Science Congress meeting in Bombay in 1938, with Fisher, who was a delegate to that Congress, as president at the Calcutta meeting. Fisher came to India for six weeks, choosing to travel by ship although passage by air was offered, mainly because of the company: the physicist Lord Rutherford, Carl Jung and two other members of the Royal Society. They sailed in November 1937 for India \cite{ref24}. 

S. S. (Subendhu Sekhar) Bose went to Bombay to accompany him by train to Calcutta after a tour through central India. Fisher delivered the Presidential address, the Governor of Bengal being present. He also intervened with the Governor and the Viceroy because Mahalanobis's sample survey of the jute crop in Bengal was being threatened to be shut down by the minister on the grounds that a small sample could not possibly have any relevance to a crop grown on millions of acres \cite{ref24}! This survey was the basis later of the National Sample Survey of India for economic and agricultural statistics, to this day crucial for a country of a billion people.

Apart from unappreciative governments and ministers, there were also disagreements between fellow statisticians. A referee of this essay has pointed to the work by V. G. Panse on the cotton crop in Madhya Pradesh and P. V. Sukhatme (who had also worked with Fisher in England) on the wheat crop in Uttar Pradesh at about the same time as Mahalanobis's on jute. They did not favour the sampling approach but instead advocated using field to field enumeration of crop yields produced by the local revenue agencies. But they also insisted on the random selection of sample plots from the revenue agency's data. Their method of `objective sampling' was extended by ICAR later to cover wheat and rice as well as other foodgrains over most of India.    

In Calcutta, Fisher discussed questions of design with the two Boses and Roy, including the large body of anthropological data available on the build and appearance of various races on the subcontinent. Mahalanobis had introduced a measure of `distance' between the races, and these discussions led later to generalized variances for distributions. Unfortunately, S. S. Bose died young the next year and the subsequent development was carried out by the others, notably R. C. Bose \cite{ref24}.

Raj Chandra Bose (see an autobiographical chapter in \cite{ref22}), born in 1901, studied mathematics at Hindu College, Delhi, moving later to Calcutta for a second M. A. and becoming a lecturer in 1930. Mahalanobis hired him in a half-time position at ISI in 1932. It is said that Bose was told one morning that a `sahab' (in Hindi: master) in a car had come to see him. This  turned out to be Mahalanobis who had seen his geometrical work and recruited him into statistics. Mahalanobis and ISI used to move to the hill station of Darjeeling in the summer months and in summer 1933, Bose was given volumes of Biometrika, a typed list of 50 papers, and Fisher's book on statistical methods as his statistical education. S. N. (Samarendra Nath) Roy \cite{ref32} was hired a few months later. 

They started working on Fisher's idea of using $n$-dimensional Euclidean space to represent $n$-samples. In 1936, F. W. Levi, who had fled from the Nazis, became head of the mathematics department at Calcutta, and they learnt from him finite fields and finite geometries. (Friedrich Wilhelm Levi later spent four years at the Tata Institute of Fundamental Research in Bombay (now Mumbai), returning in 1952 to Berlin and Freiburg where he died in 1966 \cite{ref17,ref33}.) Thus primed, Fisher's visit in 1938 and his questions on statistical designs for controlled experiments led them to use finite geometries for that. Fisher recognized the birth of a mathematical field, encouraged Bose to write up the work which was published \cite{ref12} in the {\it Annals of Eugenics} which he edited. 

In 1941, Calcutta University started a post-graduate department in statistics with Mahalanobis as head and Bose and Roy the first lecturers. Among the first batch of students was C. R. (Calyampudi Radhakrishna) Rao, another eminent Indian statistician and later himself director of the ISI. Fisher was also in India during the war (when Calcutta was in blackout) and again after, celebrating his 55th birthday in Calcutta \cite{ref24}. He returned to London for a meeting of the Royal Society where he spearheaded the election of Mahalanobis as a Fellow of the Society. Among his subsequent visits was one in 1957 for the 25th anniversary of the ISI. 

When Mahalanobis stepped down as head of the Calcutta department, Bose took the position in 1945. Later, wanting a career in research and teaching, he turned down positions with administrative duties and became a professor at the University of North Carolina in 1949. S. N. Roy joined him there the next year. Seven other Indians did their Ph.D. with Bose in that university, including S. S. Shrikhande. During his later return as a visiting professor, Shrikhande and Bose, together with E. T. Parker, disproved \cite{ref34} a 175 year old conjecture of Euler on orthogonal Latin Squares \cite{ref35}. 

Latin Squares, also related to the topics in Section II, are $s \times s$ arrangements of $s$ distinct symbols such that each occurs once in each row and column. $s$ is called the order of the square. The currently popular pastime of SuDoKu, which arranges numbers from 1 to 9 in a $9 \times 9$ square is an example of order 9. Two such squares of the same order are said to be orthogonal if, upon superposing, each symbol of one occurs exactly once with each symbol of the other. Thus, in order 2, where the only two squares possible are 
$\left( \begin{array}{cc}
0 & 1 \\
1 & 0 
\end{array} \right)$ and 
$ \left( \begin{array}{cc}
1 & 0 \\
0 & 1 
\end{array} \right)$, they are clearly not orthogonal. Upon superposition, a 0 occurs only with 1 and vice versa, never the 0-0 and 1-1 combinations. On the other hand, it is easy to construct an orthogonal pair in order 3: 
$\left( \begin{array}{ccc}
0 & 1 & 2 \\
1 & 2 & 0 \\
2 & 0 & 1 
\end{array} \right)$ and
$\left( \begin{array}{ccc}
0 & 2 & 1 \\
1 & 0 & 2 \\
2 & 1 & 0
\end{array} \right)$.

Euler conjectured that there is no pair of orthogonal Latin squares of order 6 or of order twice an odd number. 175 years later, Bose and co-workers showed that only the statement about 6 is correct but the rest of Euler's conjecture is not \cite{ref32}. Indeed, orthogonal Latin squares exist for all orders except 1, 2 and 6! The discovery led to an interview with the science editor of The New York Times and a front page story on Bose and the result. On the morning after, the hotel desk clerk recognized Bose from his photo and said, "You must have done something. The front page of The New York Times cannot be bought for a million dollars" \cite{ref22}. From 1971 to 1980, Bose was a professor at Colorado State University, then moved to emeritus status but remained active till his death in 1987. He had been elected to the U. S. National Academy of Sciences in 1976.

The connection of orthogonal Latin squares to the experimental design that Fisher was interested in is clear. A Latin square can be formed for any symbols, not necessarily numbers as in the canonical examples and in SuDoKu. Thus, consider one Latin square of potato varieties, another of fertilizer types. If they are orthogonal, every potato variety sees on it every type of fertilizer. Indeed, the existence of orthogonal Latin squares, or that of Hadamard matrices (see Appendix), is in correspondence with the existence of BIB designs. The number of Latin squares increases rapidly with the order (576 in order 4 and 161,280 in order 5). Fisher made extensive use of Latin squares in randomizing the application of treatments to varieties and produced detailed tables with Yates \cite{ref36} for this purpose. It is also interesting to note the discussion of orthogonal Latin squares of order 3 (such as in the example given above) in Fisher's book \cite{ref18} for studying the effects of nitrogen, phosphorous and potassium (the three numbers in that order on garden fertilizer bags today!) on rubber plants. Given rubber's role in war, rubber plantations in Ceylon (now Sri Lanka) and Malaya were key to the British and allied efforts during the world wars, and here again we see Fisher's very practical bent towards applied research.     

In concluding this section on the pioneering contribution of Indian mathematicians to Design Theory, their continued contributions throughout the 1930-1950s is evinced by the names already mentioned of the two Boses, Roy, Shrikhande, Rao and Savur, as well as those of D. Ray-Chaudhuri (another student of R. C. Bose with whom he started work in 1955 on coding theory), K. R. Nair (with whom Bose introduced partially balanced incomplete block designs), K. Kishen (who worked with Bose on projective geometries and so-called `factorial' designs), Q. M. Hussain, K. N. Bhattacharya and S. Chowla. 

\section{Appendix: Higher Arithmetics and other mathematical connections}

In coding theory, the so-called `packing problem' in transmitting $s$ different symbols with $t$ the measure of error-correcting capability called the `Hamming distance' and $n$ the number of redundant parity checks included in each block of transmitted symbols needs $m_t(n, s)$ as the maximum length of the block in a linear code. Fisher gave the result $m_2 (n, s)=(s^n -1)/(s-1)$ which we recognize from section IV as the size of $PG( n-1, s)$. The Indian mathematician Bose, discussed in section V, gave results for $m_3(n, 2)$, $m_3(3, s)$, and $m_3(4, s)$. See section 5, chapter XIII, volume 2 of \cite{ref1} for the use of The Fano Plane (Fig. 3) for Hamming codes.

Yet another subject with close links to designs and geometries is that of `Hadamard matrices' \cite{ref5,ref37}. Such a matrix, denoted by $H_n$, is a $n \times n$ matrix with entries $\pm 1$. $H_2$, the simplest, is
$\left( \begin{array}{cc}
1 & 1 \\
1 & -1 
\end{array} \right)$ . They can be constructed for $n$ values that are divisible by 4, and the existence of a $H_n$ implies the existence of a symmetric BIB with $(v=b=n-1, r=k=(n/2)-1, \lambda=(n/4)-1)$. $H_8$ is, therefore, associated with the $2-(7, 3, 1)$ design or the $PG (2, 2)$ Fano Plane.

Finally, as a further connection between design theory and other branches of mathematics, The Fano Plane also describes the `fourth' arithmetic. Our first acquaintance in early school is with real numbers which may be regarded as one-dimensional arithmetic (`the real line'). In high school algebra, we encounter complex numbers, $(a+ib)$, two-dimensional numbers (`the complex plane') built on reals $a$ and $b$ and the imaginary unit $i$, the square root of (-1). All the usual operations of addition, subtraction, multiplication, and division can be carried out in both cases. 

Extending further, it is well known that there is no consistent counterpart of `three-dimensional numbers', the next with all these operations being in four dimensions. Invented by Hamilton \cite{ref38} and called `quaternions', these numbers $(a+ib+jc+kd)$, built on reals $(a, b, c, d)$ and three square roots of (-1) called $(i, j, k; \,\,i^2=j^2=k^2=-1)$ provide the `third' arithmetic (more technically, a `division algebra' \cite{ref39}) upon defining the multiplication rules between these three objects. This rule is that the product of any two gives the third with a $\pm 1$ sign, depending on whether one cycles through them from left to right and then looping backwards to close the cycle, or from right to left. Thus, $(ij=k, jk=i, ki=j)$ and $(ji=-k, kj=-i, ik=-j)$. While all the four operations referred to above can be carried out with quaternions, clearly from this rule it follows that the order in which two quaternions are multiplied (or divided) matters, the multiplication not being `commutative' as in the case of reals and complex numbers. 

Quaternionic multiplication is familiar in physics, especially in quantum physics where rotation and angular motion display these anti-commutative aspects. Although less familiar, the fourth and last consistent arithmetic is that of `octonions', built similarly on seven independent square roots of (-1) \cite{ref40,ref41,ref42}. These eight-dimensional numbers involve the seven objects $e_i$ in Fig. 3, that figure providing also the multiplication rule between them. Each line has three of them and they have the anti-commutative multiplication as stated above, the product of two giving the third, with a plus sign if along the arrow and a minus sign if against. Not only is octonionic multiplication not commutative but it is not `associative' as well which means in multiplying three of them, the way they are grouped in pairs to carry out the multiplication matters. This property, familiar from reals, that $a(bc)$ and $(ab)c$ are the same, holds also for complex numbers and quaternions but fails for octonions. Not surprisingly, there is no consistent arithmetic with multiplication and division possible beyond them, and these are the only four arithmetics. (Technically, the four division algebras are distinguished by what is called `Hurwitz's' theorem, that the 'norm' of a product factorizes as the product of the norms \cite{ref39,ref40}.)

Recently, The Fano Plane of Fig. 3 has also occurred in systems of quantum spins or what are called qubits in quantum computation and quantum information \cite{ref43,ref44,ref45,ref46}. This continues the unexpected connections between various branches of mathematics and the sciences. As another human connection, The Fano Plane is named for a famous Italian geometer Gino Fano. His son, Ugo Fano, a distinguished physicist, was the doctoral father of the author of this essay.


\begin{thebibliography}{}

\bibitem{ref1} Beth T, Jungnickel D and Lenz H 1985 {\it Design Theory} (Z\"{u}rich: Bibl. Inst.) and 1993 {\it Encyclopedia of Mathematics} (Cambridge: Cambridge University Press) vol 69

\bibitem{ref2} Lenz H 1991 Half a century of Design Theory {\it Mitteilungen Math. Gesellschaft Hamburg} {\bf 12} 579-593

\bibitem{ref3} Jungnickel D 1990 Latin squares, their geometries and their groups {\it Coding Theory and Design Theory} IMA Volumes in Math. and its Appl. (Berlin: Springer)
 
\bibitem{ref4} Hall M Jr 1967 {\it Combinatorial Theory} (Waltham, MA: Blaisdell Press)

\bibitem{ref5} Rao Raghava D 1971 {\it Constructions and Combinatorial Problems in Design of Experiments} (New York: Wiley)

\bibitem{ref6} Bose R C and Manvel B 1984 {\it Introduction to Combinatorial Theory} (New York: Wiley)

\bibitem{ref7} Gropp H 1992 The birth of a mathematical theory in British India {\it Colloq. Math. Soc. Janos Bolyai} {\bf 60} 315-327

\bibitem{ref8} Woolhouse W S B 1844 Prize Question {\it Lady's and Gentleman's Diary}

\bibitem{ref9} Biggs N L 1981 T. P. Kirkman, mathematician {\it Bull. London Math. Soc.} {\bf 13} 97-120

\bibitem{ref10} Kirkman T P 1850 Query VI {\it Lady's and Gentleman's Diary} {\bf 147} 48 and Note on an unanswered prize question {\it Cambridge and Dublin Math. Journal} {\bf 5} 255-262
 
\bibitem{ref11} Yates F 1936 Incomplete randomized blocks {\it Ann. Eugenics} {\bf 7} 121-140

\bibitem{ref12} Bose R C 1939 On the construction of balanced incomplete block designs {\it Ann. Eugenics} {\bf 9} 353-399

\bibitem{ref13} Witt E 1938 \"{U}ber Steinerische Systeme {\it Abh. Hamburg} {\bf 12} 265-275

\bibitem{ref14} Steiner J 1853 Combinatorische Aufgabe {\it J. Reine Angew. Math.} {\bf 45} 181-182

\bibitem{ref15} Reiss M 1859 \'{U}ber eine Steinerische combinatorische Aufgabe {\it J. Reine Angew. Math.} {\bf 56} 326-344

\bibitem{ref16} Kirkman T P 1847 On a problem in combinations {\it Cambridge and Dublin Math. Journal} {\bf 2} 191-204

\bibitem{ref17} Gropp H 1991 The history of Steiner systems S(2,3,13) {\it Mitteilungen Math. Gesellschaft Hamburg} {\bf 12} 849-861

\bibitem{ref18} Fisher R A 1935 {\it The design of experiments} (Edinburgh: Oliver and Boyd)

\bibitem{ref19} Fisher R A 1930 {\it The genetical theory of natural selection} (Oxford: Oxford University Press)

\bibitem{ref20} Fienberg S E and Hinkley D V 1980 {\it R.A. Fisher: An Appreciation} Lecture Notes in Statistics 1 (New York: Springer)

\bibitem{ref21} Bennett J H 1983 {\it Natural selection, heredity, and eugenics} (Oxford: Clarendon Press)

\bibitem{ref22} Gani J 1982 {\it The Making of Statisticians} (Berlin: Springer-Verlag)

\bibitem{ref23} Tankard J W Jr 1984 {\it The Statistical Pioneers} (Cambridge A: Schenkman Publishing)

\bibitem{ref24} Box Joan Fisher 1978 {\it R. A. Fisher: The Life of a Scientist} (New York: John Wley Press)

\bibitem{ref25} Savur S R 1939 A note on the arrangement of incomplete blocks, when $k=3$ and $\lambda=1$ {\it Ann. Eugenics} {\bf 9} 45-49

\bibitem{ref26} Fisher R A 1940 An examination of the different possible solutions of a problem in incomplete blocks {\it Ann. Eugenics} {\bf 10} 52-75

\bibitem{ref27} Fisher R A 1941/2 New cyclic solutions to problems in incomplete blocks {\it Ann. Eugenics} {\bf 11} 290-299

\bibitem{ref28} Hirschfeld J W P 1979 {\it Projective geometries over finite fields} (Oxford: Oxford University Press)

\bibitem{ref29} Hughes D R and Piper F C 1985 {\it Design Theory} (Cambridge: Cambridge University Press)

\bibitem{ref30} Bennett C H and Brassard 1984 {\it Proceedings of the IEEE Conference on Computers, Systems and Signal Processing, Bangalore, India} (New York: IEEE) 175-179

\bibitem{ref31} Ekert A 1991 Quantum cryptography based on Bell's theorem {\it Phys. Rev. Lett.} {\bf 67} 661-663

\bibitem{ref32} http://en.wikipedia.org/wiki/S\_N\_Roy

\bibitem{ref33} Pinl M 1971/2 Kollegen in einer dunklen Zeit, III Teil {\it Jahresbericht der Deutschen Mathematiker-Vereinigung} {\bf 73} 153-208

\bibitem{ref34} Bose R C, Parker E T and Shrikhande S 1960 On orthogonal Latin squares {\it Can. J. Math.} {\bf 12} 189-203

\bibitem{ref35} Biggs N L 1985 {\it Discrete Mathematics} (Oxford: Clarendon Press)

\bibitem{ref36} Fisher R A and Yates F 1949 {\it Statistical tables for biological, agricultural and medical research} (London: Oliver and Boyd) 3rd ed 

\bibitem{ref37} Hadamard J 1893 Resolution d'une question relative aux determinants {\it Bull. Sci. Math.} {\bf 2} 240-246

\bibitem{ref38} Hankins T L 1980 {\it Sir William Rowan Hamilton} (Baltimore: Johns Hopkins University Press)

\bibitem{ref39} Dickson L E 1919 On Quaternions and Their Generalization and the History of the Eight Square Theorem  {\it Ann. Math.} {\bf 20} 155-171

\bibitem{ref40} Dixon G M 1994 {\it Division Algebras: Octonions, Quaternions, Complex Numbers and the Algebraic Design of Physics} Mathematics and its Applications, Vol. 290 (Dordrecht: Kluwer Press)

\bibitem{ref41} Coxeter H S M 1946 Integral Cayley Numbers {\it Duke Math. Journal} {\bf 13} 561-578

\bibitem{ref42} Baez J C 2001 The Octonions {\it Bull. New Ser., Am. Math. Soc.} {\bf 39} 145-205 and www.jmath.usr.edu/home/baez/octonions

\bibitem{ref43} Rau A R P 2009 Mapping two-qubit operators onto projective geometries {\it Phys. Rev. A} {\bf 79} 042323 (1-6)

\bibitem{ref44} Planat M and Saniga M 2008 On the Pauli graphs on $N$-qudits {\it Quantum Inf. Comput.} {\bf 8} 127-146

\bibitem{ref45} Levay P, Saniga M and Vrana P 2008 Three-Qubit Operators, the Split Cayley Hexagon of Order Two and Black Holes {\it Phys. Rev. D} {\bf 78} 124002 (1-22)

\bibitem{ref46} Rau A R P 2009 Algebraic characterization of $X$-states in quantum information arXiv:0906.4716 and {\it J. Phys. A: Math. Gen.} {\bf 42}, 412002 (1-7)

\end{thebibliography}
\end{document}